\documentclass[12pt]{article}
\usepackage{amsmath,amssymb,graphicx}
\usepackage{graphics,epsf}

\usepackage{subfloat}
\newcommand{\be}{\begin{equation}}
\newcommand{\bea}{\begin{eqnarray}}
\newcommand{\eea}{\end{eqnarray}}
\newcommand{\ba}{\begin{array}}
\newcommand{\ea}{\end{array}}
\newcommand{\ee}{\end{equation}}

\begin{document}

\begin{titlepage}

\vspace*{5mm}%
\begin{center}

{{\Large {\bf Holographic insulator/superconductor phase transition in Born-Infeld electrodynamics}}}

\vspace*{15mm} \vspace*{1mm} {Nan Bai$^{a}$, Yi-Hong Gao$^{a}$, Bu-Guan Qi$^{a}$ and Xiao-Bao Xu$^{a}$}

 \vspace*{2cm}

{\it ${}^a$ State Key Laboratory of Theoretical Physics,\\
Institute of Theoretical Physics,\\ Chinese Academy of
Sciences, P.O. Box 2735, Beijing 100190, China }

\vspace*{.4cm}

 bainan@itp.ac.cn, guanbu@itp.ac.cn, gaoyh@itp.ac.cn, and xbxu@itp.ac.cn

\vspace*{2cm}
%%\maketitle
\end{center}
\begin{abstract}
We studied holographic insulator/superconductor phase transition in the framework of Born-Infeld electrodynamics both numerically and analytically.
First we numerically study the effects of the Born-Infeld electrodynamics on the phase transition, find that the critical chemical potential is not changed by the Born-Infeld parameter. Then we employ the variational method for the Sturm-Liouville eigenvalue problem to analytically study the phase transition. The analytical results obtained are found to be consistent with the numerical results.
\end{abstract}
\end{titlepage}

\section{Introduction}

The AdS/CFT correspondence \cite{{adscft1},{adscft2},{adscft3}} which relates string theory on asymptotically AdS spacetime to a conformal field theory on the boundary, has been extensively applied in nuclear physics and condensed matter physics. In particular the related developments of dual gravitational models which  describe insulator/superconductor phase transitions in condensed matter physics have been proposed in \cite{{hs1},{hs2}}.

The Einstein-Maxwell theory coupled to a charged scalar field would be the simplest model for holographic superconductor. In this model, there exists a critical temperature $T_c$ \cite{hs1}, below which charged AdS black holes may have non-zero scalar hair. In the context of AdS/CFT connection, a charged scalar field is dual to an operator which carries the global $U (1)$ charge. Having non-zero scalar hair indicates that the expectation value of dual operator is non-zero which leads to the break of global $U(1)$ symmetry, thus can be understood as a second order conductor/superconductor phase transition. The behavior of AC conductivity in these phases \cite{hs2} make  the interpretation above reliable.

There have been many discussions concerning holographic superconductor models among which general relativity coupled to Maxwell field plus a charged
scalar field attracts most attention \cite{hs3}-\cite{n1} and the related Lagrangian is
\be
 \mathcal{L}=-\frac{1}{4}F^2-|\nabla_{\mu}\psi-iq A_{\mu}\psi|^2-m^2|\psi|^2,\qquad F=F_{\mu\nu}F^{\mu\nu}
\ee
However,the non-linear extension of the original Maxwell electrodynamics to Born-Infeld (BI) model \cite{BI} also caused intensive investigations \cite{n2}-\cite{hs20}, since BI electrodynamics possesses many interesting physical properties as finite total energy and the invariance under electromagnetic duality. In string theory, BI action can also describe gauge fields which arise from D-brane attached with open strings \cite{{Gibbons},{Tseytlin}}. The effects of BI electrodynamics on the holographic superconductors in the background of Ads-Schwarzschild black hole spacetime has been analyzed numerically in \cite{hs15} and analytically in \cite{hs18}.The results shows it is much more difficult to have scalar condensation in BI electrodynamics which is similar to the one of holographic superconductors in Einstein-Gauss-Bonnet gravity \cite{{hs5},{hs12}}, where the higher curvature corrections make condensation harder.

The authors of \cite{hs7} have constructed a model describing an insulator/superconductor phase transition at zero temperature, using a five dimensional AdS soliton in an Einstein-Maxwell-charged scalar field. The generalization of holographic insulator/superconductor has been carried out \cite{{Insulator1},{Insulator2},{Insulator3}}.The aim of the present article is to extend holographic insulator/superconductor in the framework of Born-Infeld electrodynamics. The main purpose here is to see what effect of Born-Infeld scale parameter on the holographic insulator/superconductor. With both numerical and analytic method, we find a bit unexpected result that the effect of Born-Infeld electrodynamics is too small to be seen in our study, but it is consistent with \cite{hs20}, which state that ``the critical chemical potential is not changed by the coupling parameter $b$".

The paper is organized as follows. In section 2 we study holographic insulator/super-conductor in the framework of Born-Infeld electrodynamics in the probe limit. We numerically calculate the critical chemical potential $\mu_c$, the condensations of the scalar operators $\langle \cal{O}\rangle$ and charge density $\rho$. We also study the conductivity of the theory. In section 3 we reproduce the same result as the numerical one for the critical chemical potential $\mu_c$ by the Sturm-Liouville analytical method. The last section is devoted  to conclusions.

%%%%%%%%%%%%%%%%%%%%%%%%%%%%%%%%%%%%%%%%%%%%%%%%%%%%%%%%%%%%%
%%%%%%%%%%%%%%%%%%%%%%%%%%%%%%%%%%%%%%%%%%%%%%%%%%%%%%%%%%%%%

\section{Numerical analysis of the critical chemical potential $\mu_c$}
In order to construct the holographic insulator/superconductor phase transition, we consider the fixed
background of five dimensional AdS soliton, e.g. in the probe limit.
The Ads soliton metric reads
\begin{align}
 ds^2&= L^2\frac{dr^2}{f(r)} + r^2( -dt^2 + dx^2 + dy^2 ) +
 f(r)d\chi^2 \ ,\\
  f(r) &= r^2 - \frac{r_0^4}{r^2}
\end{align}
where $L$ is the AdS radius and $r_0$  the tip of the soliton which is a conical singularity in this solution. Essentially this solution can be obtained from a five dimensional AdS black hole solution by making use of double Wick rotations \cite{hs7}. We can also set $L = 1$ and $r_0$ = 1 without the loss of generality.

We now consider a charged complex scalar field and an electric field  in the above background, we investigate Born-Infeld electrodynamics instead of Maxwell's therefore the corresponding Lagrangian density can be
\begin{eqnarray}
\mathcal{L}=\mathcal{L}_{BI}-
|\nabla_{\mu}\psi-iqA_{\mu}\psi|^2-m^2|\psi|^2
\label{m10}
\end{eqnarray}
where $\psi$ is a charged complex scalar field, $\mathcal{L}_{BI}$ is the Lagrangian density of the Born-Infeld electrodynamics
\begin{eqnarray}
\mathcal{L}_{BI}=\frac{1}{b}\bigg(1-\sqrt{1+\frac{b F}{2}}\bigg). \quad F\equiv F_{\mu\nu}F^{\mu\nu}
\label{m11}
\end{eqnarray}
 The BI coupling parameter $b$ describes the nonlinearity of electrodynamics. The above BI lagrangian could be reduced to Maxwell lagrangian $-\frac{1}{4}F_{\mu\nu}F^{\mu\nu}$ naturally in the limit $b\rightarrow 0$ . We further assume that the BI coupling parameter $b$ is small,since the Ads Soliton of background spacetimes is the solution of the Einstein equation in Einstein-Maxwell-scalar field theory.

To study the insulator/superconductor phase transition we will consider the following ansatz \cite{hs7}
\begin{eqnarray}
A_{\mu}=(\phi(r),0,0,0),\;\;\;\;\psi=\psi(r)
\label{vector}
\end{eqnarray}
The equations of motion are
\begin{eqnarray}
\psi''+\left(\frac{f'}{f}+\frac{3}{r}\right)\psi'
+\left(\frac{q^2\phi^{2}}{r^2f}-\frac{m^2}{f}\right)\psi=0 \label{e1}
\end{eqnarray}
and
\begin{eqnarray}
\phi''+\left(\frac{1}{r}+\frac{f'}{f}\right)\phi'-\left(\frac{bf'}{2r^2}+\frac{2bf}{r^3}\right)\phi'^3
-\frac{2q^2\psi^2}{f}\bigg(1-\frac{bf}{r^2}\phi'^2\bigg)^{3/2}\phi=0 \label{e2}
\end{eqnarray}
To seek the solutions of the equations (\ref{e1}) and (\ref{e2}), we impose the boundary condition at the tip $r=r_0$ and the boundary
$r=\infty$. Near the boundary, the scalar field $\psi(r)$ and the scalar potential $\phi(r)$ behave as
\begin{align}
 \psi& = \frac{\psi_{-}}{r^{\lambda_{-}}} + \frac{\psi_{+}}{r^{\lambda_{+}}}\label{f3} ,\\
 \phi& = \mu - \frac{\rho}{r^2}\label{f4}
\end{align}
where
\begin{eqnarray}
\lambda_{\pm}=2\pm\sqrt{4+m^2}
\end{eqnarray}
is the conformal dimension of the condensation operator $\langle \cal{O}\rangle$ belongs to the boundary field theory,and
$\psi_{-}$ and $\psi_{+}$ are the vacuum expectation values for the boundary
operator $\langle \cal{O}\rangle$. $\mu$ and $\rho$ are the chemical potential and the charge density,respectively.

For simplicity we will take $m^2=-\frac{15}{4}$, which is above the Breitenlohner-Freedman bound \cite{bf2}, hence $\psi  = \frac{\psi_{-}}{r^{\frac{3}{2}}} + \frac{\psi_{+}}{r^{\frac{5}{2}}}$. In order to have a stable boundary theory, we impose the boundary condition as $\psi_{-}=0$ or $\psi_{+}=0$ \cite{hs2}.
In this paper, we shall set $\psi_{-}=0$ and $\langle \cal{O}\rangle = \psi_{+}$.

\noindent{}At the tip, we have
\begin{align}
 \psi&= \alpha_0 + \alpha_1\log (r-r_0) + \alpha_2(r-r_0) + \dots \label{f1} ,\\
 \phi&= \beta_0 + \beta_1\log (r-r_0) + \beta_2(r-r_0) + \dots \label{f2}
\end{align}
We can impose the Neumann boundary condition $\alpha_1=\beta_1=0$ to make the physical
quantities finite. Note that the equations of motion (\ref{e1}) and (\ref{e2}) have the scaling symmetry, $(\psi,\phi,\mu,q,b)\to (\lambda\psi,\lambda\phi,\lambda\mu,q/\lambda,b\lambda^2)$. Thus we will set $q=1$ in the following discussions. Introducing a new variable $z=1/r$, we rewrite the equation of motion (\ref{e1}) and (\ref{e2}) into
\begin{eqnarray}
 \psi''+\left(\frac{f'}{f}-\frac{1}{z}\right)\psi'+\left(\frac{\phi^{2}}{z^2f}-
 \frac{m^2}{z^4f}\right)\psi=0 \label{e3}
\end{eqnarray}
and
\begin{eqnarray}
 \phi''+\left(\frac{f'}{f}+\frac{1}{z}+\left(2f-\frac{zf'}{2}\right)bz^5\phi'^2\right)\phi'
 -\frac{2\psi^2}{z^4f}\left(1-bfz^6\phi'^2\right)^{3/2}\phi=0 \label{e4}
\end{eqnarray}

We use shooting method numerically calculate the critical potential $\mu_c$, the condensations of the scalar operators $\langle \cal{O}\rangle$ and charge density $\rho$ for $b=0.01$. Then we change Born-Infeld parameter $b$ of the theory, choosing $b=0,  0.1,  0.2, 0.3$, find that the result varied so small that we can neglect the effect of Born-Infeld parameter $b$.  It is a somewhat surprising result to us. However, our analytic calculation support this.
\begin{figure}
\begin{center}
%\vspace{1cm}

\includegraphics[height=4.86cm, width=8cm]{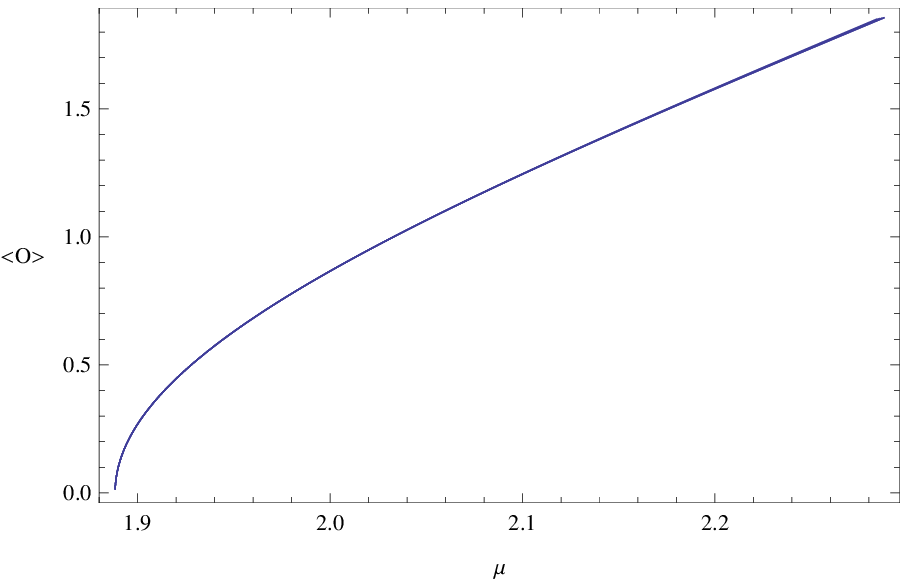}
\includegraphics[height=5cm, width=7.6cm]{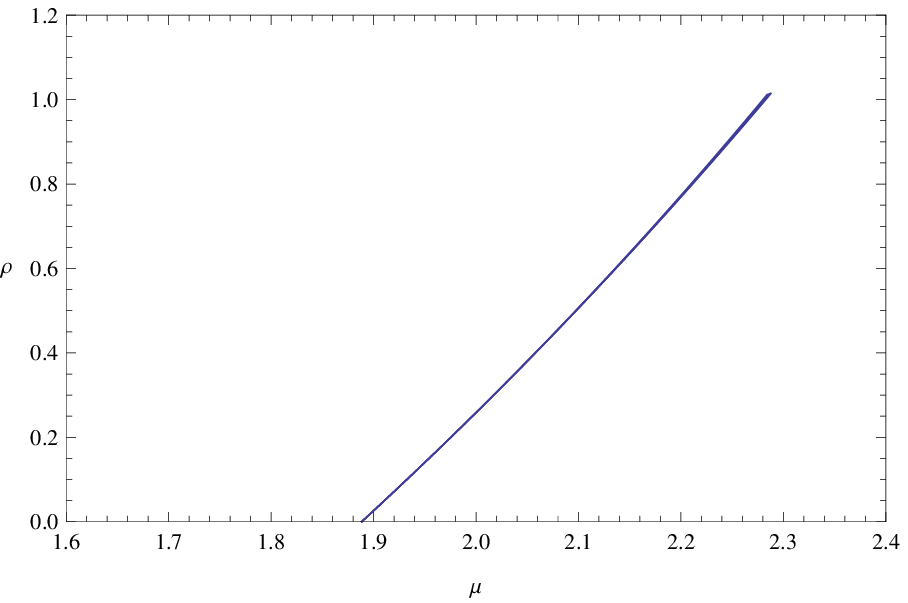}
\caption{ Here we plot the behaviors of $\langle \cal{O}\rangle$  and charge density $\rho$
for various values of chemical potential $\mu$. Here $\mu_c=1.888$.} \label{fig1}
\end{center}
%\vspace{0.5cm}
\end{figure}

Now we holographically calculate the conductivity  $\sigma(\omega)$ by perturbing the gauge field \cite{hs2}. To do so, we consider the following ansatz for the gauge field along $x$ direction
\be
A_x=A_x(r)e^{-i\omega t}
\ee
Neglecting the backreaction of the perturbational field on background metric, one get the equation of motion
of $A_x$ :
\begin{align}
\begin{split}
A_x'' + \left( \frac{f'}{f} + \frac{1}{r} \right)A_x' + \left(
 \frac{\omega^2}{r^2 f} - \frac{2\psi^2}{f}\big[1+b(-\frac{f\phi'^2}{r^2}-\frac{\omega^2 A_x^2}{r^4}+\frac{f A_x'^2}{r^2})\big]^{1/2}\right) A_x\\
 -\frac{b}{2f}\frac{1}{1+b(-\frac{f\phi'^2}{r^2}-\frac{\omega^2 A_x^2}{r^4}+\frac{f A_x'^2}{r^2})}\big[\frac{2f\phi'^2}{r^3}-\frac{f'\phi'^2}{f}-\frac{2f\phi'\phi''}{r^2}+\frac{4\omega^2A_x^2}{r^5}-\\
 \frac{2\omega^2A_xA_x'}{r^4}-\frac{2fA_x'^2}{r^3}+\frac{f'A_x'^2}{r^2}+\frac{2fA_x'A_x''}{r^2}\big]=0  \label{axeom}
\end{split}
\end{align}
Near the boundary($r\to \infty$), the behavior of the gauge field is
\be
A_x=A_x^{(0)}+\frac{A_x^{(1)}}{r^{2}}+\frac{A_x^{(0)}\omega^2}{2}\frac{\log\Lambda r}{r^2}+\cdots
\ee
the holographic conductivity is given as follows \cite{hs8}
\be
\sigma(\omega)=\frac{-2iA_x^{(1)}}{\omega A_x^{(0)}} + \frac{i\omega}{2}
\ee

\begin{figure}
\begin{center}
\includegraphics[width=7cm]{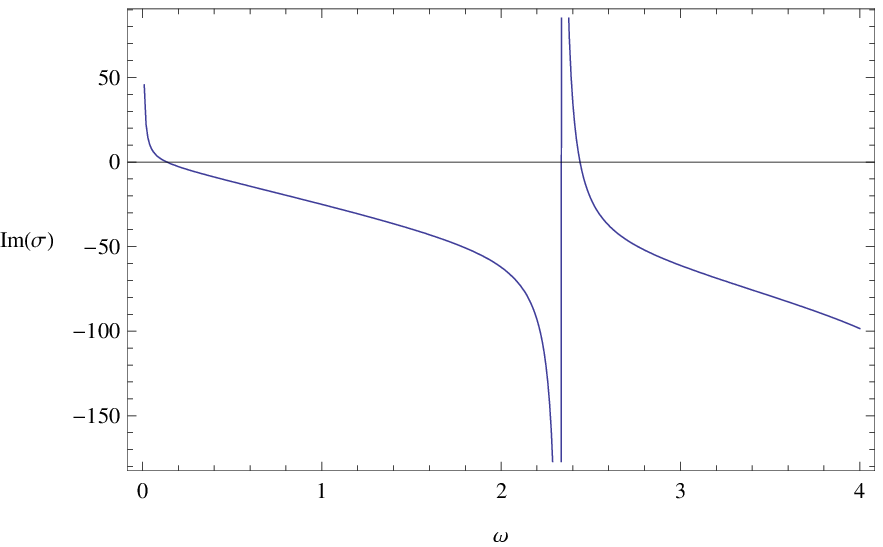}\;\;\;\;
\includegraphics[width=7cm]{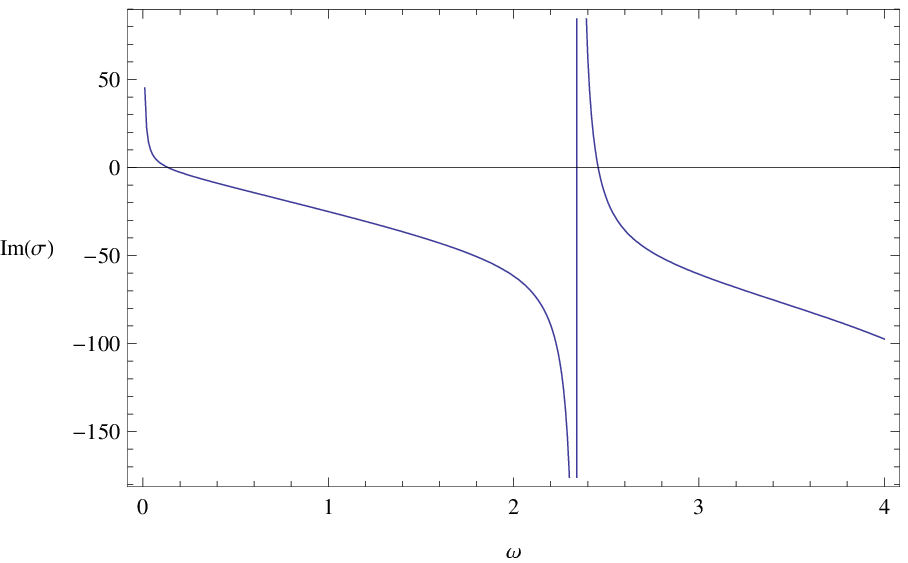}\\
\includegraphics[width=7cm]{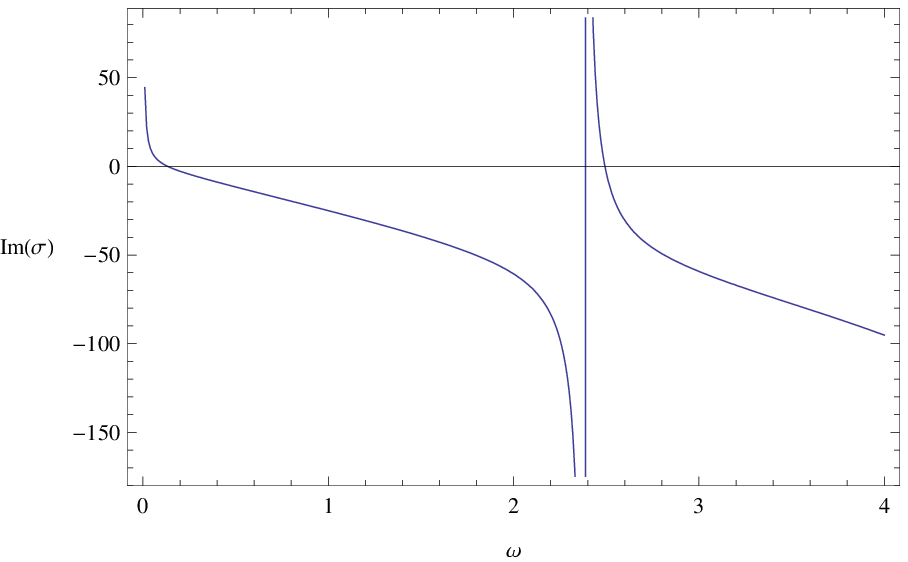}\;\;\;\;
\includegraphics[width=7cm]{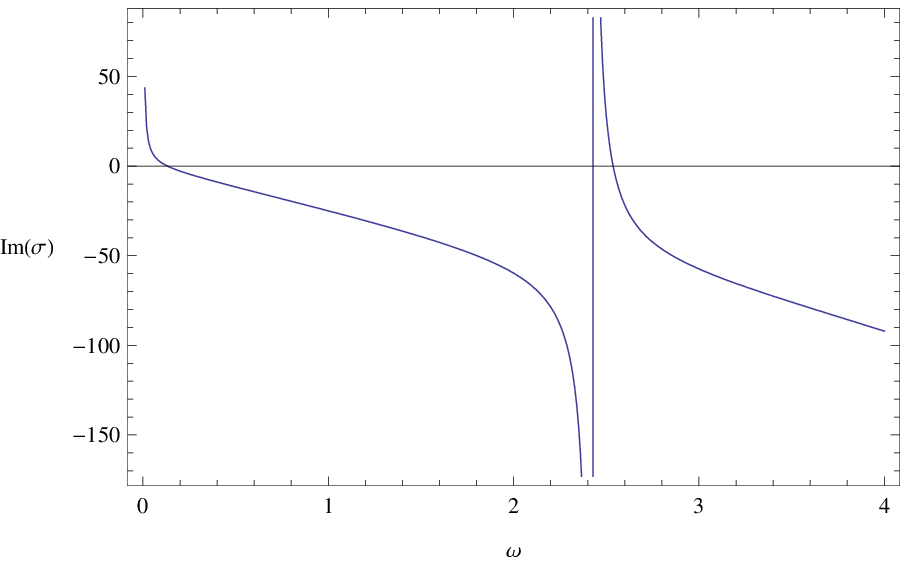}
\caption{The conductivity for operator $\langle\mathcal{O}\rangle$
for different values of $b$.    }
\end{center}
\end{figure}

In the figure 2 we have plotted the behavior of the conductivity for different Born-Infield parameter $b$
when the condensation is non-zero, we find a pole at $\omega\to 0$ suggesting that we have an infinite
conductivity according to the Kramers-Kronig relation as expected for the superconductor phase. We observed the effect of Born-Infield parameter $b$
on the conductivity, the gap frequency becomes larger as Born-Infield parameter $b$ increases. This is different from those of Born-Infeld electrodynamics in
Schwarzschild AdS black hole spacetime\cite{hs15}.

\section{Analytical calculation of the critical chemical potential $\mu_c$ }
From the results of the numerical calculations above, we found there exists a critical chemical potential $\mu_c$. When chemical potential $\mu$ exceeds this critical value,the solution is unstable and a scalar hair will emerge which means the condensations take place.This is the superconductor phase.For $\mu<\mu_c$,the scalar field is zero,which can be viewed as an insulator phase.The critical chemical potential $\mu_c$ can be obtained by using Sturm-Liouville (SL) eigenvalue method.

When $\mu\leq\mu_c$,the scalar field $\psi=0$,Eq.(\ref{e4}) reduces to
\begin{eqnarray}
 \phi''+\left(\frac{f'}{f}+\frac{1}{z}+\left(2f-\frac{zf'}{2}\right)bz^5\phi'^2\right)\phi'=0\label{e6}
\end{eqnarray}
By substituting $\phi'$ to a new variable ,the equation will be transformed into a first order ODE of Bernoulli type, see  appendix A  , and the general solution is
\begin{eqnarray}
\begin{split}
\phi\left(z\right)=&\frac{b^2 \left(117 z^8-306 z^4+221\right) z^9}{1989}-\frac{2 b C_1 \left(77 z^{12}-315 z^8+495 z^4-385\right) z^3}{1155}+\\
&C_1^2
   \left(\frac{z^{13}}{13}-\frac{4 z^9}{9}+\frac{6 z^5}{5}-\frac{1}{3 z^3}-4 z\right)+C_2\\
\end{split}
\end{eqnarray}
where $C_1$ and $C_2$ are the integration constants. The boundary condition (\ref{f4}) near the infinity $z=0$  imposes $C_1=0$ which keeps $\phi$ a finite value and $C_2=\mu$.
Thus,
\begin{eqnarray}
 \phi\left(z\right)=\frac{b^2 \left(117 z^8-306 z^4+221\right) z^9}{1989}+\mu\label{e5}
\end{eqnarray}
When $\mu\rightarrow\mu_c$ ,we take  $\phi\left(z\right)$ above into the equation (\ref{e3}),thus the equation of motion for $\psi$ becomes
\begin{eqnarray}
 \psi (z) \left(\frac{\left(\frac{b^2 \left(117 z^8-306 z^4+221\right) z^9}{1989}+\mu
   \right)^2}{z^2 f(z)}-\frac{m^2}{z^4
   f(z)}\right)+\left(\frac{f'(z)}{f(z)}-\frac{1}{z}\right) \psi '(z)+\psi ''(z)=0
\end{eqnarray}
As in \cite{hs6},we introduce a trial function $F\left(z\right)$ near the boundary $z=0$ which satisfies
\begin{eqnarray}
 \psi(z)\sim \langle{\cal O}_{i}\rangle z^{\lambda_i}F(z),
\end{eqnarray}
 We impose the boundary condition for $F(z)$ as $F(0)=1$ and $F'(0)=0$.Thus the equation of motion for $F(z)$ is
\begin{eqnarray}
\begin{split}
&F''(z)+\left(\frac{f'(z)}{f(z)}+\frac{2}{z}\right) F'(z)+\\
& \left(\left(B^2 \left(\frac{z^{17}}{17}-\frac{2 z^{13}}{13}+\frac{z^9}{9}\right)+\mu \right)^2+\frac{3}{2} \left(z f'(z)-f(z)\right)+\frac{3
   f(z)}{4}-\frac{m^2}{z^2}\right)\frac{F(z)}{z^2 f(z)}=0\\
\end{split}
\end{eqnarray}
Following the standard procedure of Sturm-Liouville eigenvalue problem \cite{ref2},we introduce a new function $T(z)$
\begin{eqnarray}
 T(z)=\left(1-z^4\right) z^{2 \lambda_i -3}
\end{eqnarray}
Multiplying $T(z)$ to both sides of the equation of $F(z)$ ,we have
\begin{eqnarray}
\begin{split}
&\left(T(z) F'(z)\right)'+\frac{F(z) T(z)} {1-z^4}\left(\left(b^2 \left(\frac{z^{17}}{17}-\frac{2 z^{13}}{13}+\frac{z^9}{9}\right)+\mu
   \right)^2-\frac{m^2}{z^2}+\right.\\
   &\left.(\lambda_i -1) \lambda_i  \left(\frac{1}{z^2}-z^2\right)+\lambda_i  \left(\left(-\frac{2}{z^3}-2 z\right)
   z+z^2-\frac{1}{z^2}\right)\right)=0
\end{split}
\end{eqnarray}
Furthermore we define another two functions $U$ and $V$ as
\begin{eqnarray}
\begin{split}
 U=&\frac{1}{1-z^4}\left(b^4 \left(\frac{z^{17}}{17}-\frac{2 z^{13}}{13}+\frac{z^9}{9}\right)^2+2 b^2 \mu  \left(\frac{z^{17}}{17}-\frac{2 z^{13}}{13}+\frac{z^9}{9}\right)-\frac{m^2}{z^2}+\right.\\
 &\left.(\lambda_i
   -1) \lambda_i  \left(\frac{1}{z^2}-z^2\right)+\lambda_i  \left(\left(-\frac{2}{z^3}-2 z\right) z+z^2-\frac{1}{z^2}\right)\right)
\end{split}
\end{eqnarray}
\begin{eqnarray}
 V=T(z)/(1-z^4)
\end{eqnarray}
Then we can obtain the eigenvalue of $\mu^2$ by substituting $U$ and $V$ into the expression
\begin{eqnarray}\label{eigenvalue}
\mu^{2}=\frac{\int^{1}_{0}T\left(F'^{2}-UF^{2}\right)dz}{\int^{1}_{0}VF^{2}dz},\label{b2}
\end{eqnarray}
In order to estimate the minimum value of $\mu^2$,we assume the trial function to be $F(z)=1-az^2$,where $a$ is a constant.In the normal case ,the r.h.s of $\mu^2$ is an expression of $a$ and other parameters already given as constants, e.g  Born-Infeld parameter b in our case. But the problem we are dealing with is a bit different for the expression of $\mu^2$ contains $\mu$ as well,
\begin{eqnarray}
 \mu^2=\mu^2\left(a,b,\mu\right)
\end{eqnarray}
To avoid this difficulty ,we suppose $\mu$ on the r.h.s of the above expression to be $\tilde{\mu}$ at first, and consider $\tilde{\mu}$ as a new constant parameter. Then we can compute the extreme value of $\mu^2$ as we have done in the normal situation. But at this time the values of $a$ corresponding to the extremal eigenvalues of $\mu^2$ are related to the parameter $\tilde{\mu}$, therefore,
\begin{eqnarray}
 \mu^2=\mu^2\left(a(\tilde{\mu}),b,\tilde{\mu}\right)=\mu^2(\tilde{\mu},b)
\end{eqnarray}
At last ,for consistency, $\tilde{\mu}$ is required to be $\mu$,and we obtain an equation for $\mu$
\begin{eqnarray}
 \mu_m^2=\mu_m^2(\mu_m,b)
\end{eqnarray}
where a subscript ``m'' is added to indicate this is an equation of extreme value $\mu$. For the calculation details, see appendix B.
The analytical calculation shows the minimum eigenvalues of $\mu^2$ and the corresponding values of $a$ for different Born-Infeld parameter factor $b$ are almost the same,which is $a=0.330$ and $\mu_{min}=1.890$. We see the analytical results fit perfectly with the numerical ones.
\section{Conclusion}
In this paper ,we considered the Born-Infeld electrodynamics in the AdS soliton background.By numerical calculations , we found there exists a critical chemical potential $\mu_c$.When the chemical potential $\mu$ is lower then the critical chemical potential $\mu_c$,the AdS soliton background is stable and dual field theory can be viewed as an insulator.When $\mu>\mu_c$ ,the background begins to be unstable and a scalar hair will be developed as the condensation, thus can be interpreted as the superconductor phase.So the insulator and superconductor phase transition appeared in the Maxwell electrodynamics also exists in its non-linear extension.We compute the critical chemical potential $\mu_c$ both analytically and numerically and the results of these two methods fit perfectly good.We also found the effect of different Born-Infeld parameters are very small that even can be neglected.

\section*{Acknowledgments}

We are very grateful of M. Alishahiha and L. Li for their  useful helps.
\appendix

\section{Bernoulli type first order ODE}
The standard Bernoulli type ordinary equation takes the form
\begin{eqnarray}
 y'=f(t,y(t))=g(t)y(t)+h(t){y(t)}^n &&n\neq1 \label{b1}
\end{eqnarray}
We make the substitution $\xi=\phi'$ in the equation (\ref{e6}) which reduces to
\begin{eqnarray}
 \frac{d\xi}{dz}=-\left(\frac{f'}{f}+\frac{1}{z}\right)\xi-\left(2f-\frac{zf'}{2}\right)bz^5\xi^3
\end{eqnarray}
It is a Bernoulli type ordinary equation and the idea to solve the equation (\ref{b1}) is to convert it into a linear ODE.
Make the substitution
\begin{eqnarray}
 v(t)={y(t)}^{1-n}
\end{eqnarray}
thus
\begin{eqnarray}
 v(t)'=(1-n){y(t)}^{-n}y(t)'
\end{eqnarray}
\begin{eqnarray}
 y(t)'=\frac{y(t)^n}{1-n}v(t)'
\end{eqnarray}
so
\begin{eqnarray}
 v(t)'=(1-n)g(t)v(t)+(1-n)h(t)
\end{eqnarray}
which is a standard linear ODE,and can be solved by the standard procedure.

\section{Computation of $\mu_{min}$}
Substituting $U$,$V$ into expression (\ref{b2}) ,we get

\begin{eqnarray*}
\mu ^2=\frac{A(a)\tilde{\mu}+B(a)}{C(a)}
\end{eqnarray*}
\begin{eqnarray*}
A(a)&=&-254592 b^2 \lambda_i  \left(\lambda_i ^{14}+141 \lambda_i ^{13}+8953 \lambda_i ^{12}+337449 \lambda_i ^{11}+8370439 \lambda_i
   ^{10}+\right.\\
   & &\left.142959663 \lambda_i ^9+1704340979 \lambda_i ^8+13997457507 \lambda_i ^7+75168485084 \lambda_i ^6+\right.\\
   & &\left.226063117896 \lambda_i
   ^5+135629702768 \lambda_i ^4-1262172448656 \lambda_i ^3-\right.\\
   & &\left.2753135394624 \lambda_i ^2+1021968576000 \lambda_i +2540624486400\right)
   \left(a^2 \left(64 \lambda_i ^6+\right.\right.\\
   & &\left.\left.2848 \lambda_i ^5+53040 \lambda_i ^4+517088 \lambda_i ^3+2740052 \lambda_i ^2+7415574 \lambda_i
   +7977879\right)-\right.\\
   & &\left.2 a \left(64 \lambda_i ^6+2912 \lambda_i ^5+55888 \lambda_i ^4+570128 \lambda_i ^3+3209404 \lambda_i ^2+9320246 \lambda_i
   +\right.\right.\\
   & &\left.\left.10774995\right)+64 \lambda_i ^6+2976 \lambda_i ^5+58864 \lambda_i ^4+628992 \lambda_i ^3+3790660 \lambda_i ^2+\right.\\
   & &\left.12084582 \lambda_i
   +15758847\right)
\end{eqnarray*}
\begin{eqnarray*}
C(a)&=&3956121 (\lambda_i -2) (\lambda_i -1) \lambda_i  (\lambda_i +1) (\lambda_i +8) (\lambda_i +9) (\lambda_i +10) (\lambda_i +11) (\lambda_i +12)\\& &
   (\lambda_i +13) (\lambda_i +14) (\lambda_i +15) (\lambda_i +16) (\lambda_i +17) (\lambda_i +18) (2 \lambda_i +7) (2 \lambda_i +9)\\& & (2 \lambda_i
   +11) (2 \lambda_i +13) (2 \lambda_i +15) (2 \lambda_i +17) (2 \lambda_i +19) \left(\frac{a^2}{\lambda_i +1}-\frac{2 a}{\lambda_i
   }+\frac{1}{\lambda_i -1}\right)
\end{eqnarray*}

\begin{eqnarray*}
B(a)&=&a^2 \left(8 \lambda_i ^3+180 \lambda_i ^2+1318 \lambda_i +3135\right) \left(16 \lambda_i ^{13}+1504 \lambda_i ^{12}+61872 \lambda_i
   ^{11}+\right.\\
   & &\left.1461236 \lambda_i ^{10}+21801935 \lambda_i ^9+212273937 \lambda_i ^8+1333101101 \lambda_i ^7+\right.\\
   & &\left.4996108013 \lambda_i ^6+8010950487
   \lambda_i ^5-11997217421 \lambda_i ^4-68501188673 \lambda_i ^3-\right.\\
   & &\left.66321743349 \lambda_i ^2+59135273262 \lambda_i +73109116080\right)
   \left(2 \left(-8 \left(64 b^4-\right.\right.\right.\\
   & &\left.\left.\left.35605089\right) \lambda_i ^5-96 \left(368 b^4-85715955\right) \lambda_i ^4-16 \left(59872
   b^4-\right.\right.\right.\\
   & &\left.\left.\left.7536410505\right) \lambda_i ^3-48 \left(258472 b^4-19251803493\right) \lambda_i ^2-\right.\right.\\
   & &\left.\left.1728 \left(39269 b^4-1933664031\right)
   \lambda_i +3956121 \lambda_i ^6+3828259169280\right)+\right.\\
   & &\left.3956121 \left(\lambda_i ^5+70 \lambda_i ^4+1940 \lambda_i ^3+26600 \lambda_i
   ^2+180384 \lambda_i +483840\right) m^2\right)-\\
   & &2 a \lambda_i  \left(16 \lambda_i ^{11}+1632 \lambda_i ^{10}+73016 \lambda_i ^9+1880616
   \lambda_i ^8+30716889 \lambda_i ^7+\right.\\
   & &\left.329379372 \lambda_i ^6+2305033064 \lambda_i ^5+9928719744 \lambda_i ^4+21299033680 \lambda_i
   ^3-\right.\\
   & &\left.4869591744 \lambda_i ^2-119590571520 \lambda_i -169885900800\right) \left(3956121 \left(8 \lambda_i ^8+\right.\right.\\
   & &\left.\left.676 \lambda_i ^7+24482
   \lambda_i ^6+495859 \lambda_i ^5+6136697 \lambda_i ^4+47461414 \lambda_i ^3+\right.\right.\\
   & &\left.\left.223683648 \lambda_i ^2+586275651 \lambda_i +652759965\right)
   \left(2 \lambda_i  (\lambda_i +2)+(\lambda_i +1) m^2\right)-\right.\\
   & &\left.128 b^4 \left(64 \lambda_i ^9+5408 \lambda_i ^8+195792 \lambda_i ^7+3929640
   \lambda_i ^6+46829160 \lambda_i ^5+\right.\right.\\
   & &\left.\left.322610184 \lambda_i ^4+1147216028 \lambda_i ^3+1425431605 \lambda_i ^2-1194241044 \lambda_i
   -\right.\right.\\
   & &\left.\left.1751976837\right)\right)+\lambda_i  \left(16 \lambda_i ^{12}+1792 \lambda_i ^{11}+88864 \lambda_i ^{10}+2567628 \lambda_i ^9+\right.\\
   & &\left.47829603
   \lambda_i ^8+599455521 \lambda_i ^7+5108576957 \lambda_i ^6+29104930589 \lambda_i ^5+\right.\\
   & &\left.104893992911 \lambda_i ^4+206701115187 \lambda_i
   ^3+113193419679 \lambda_i ^2-\right.\\
   & &\left.236408070717 \lambda_i -223243908030\right) \left(3956121 \left(8 \lambda_i ^8+612 \lambda_i ^7+19974
   \lambda_i ^6+\right.\right.\\
   & &\left.\left.362715 \lambda_i ^5+4001532 \lambda_i ^4+27406788 \lambda_i ^3+113514256 \lambda_i ^2+\right.\right.\\
   & &\left.\left.259082880 \lambda_i +248371200\right)
   \left(2 \lambda_i +m^2\right)-128 b^4 \left(64 \lambda_i ^8+4832 \lambda_i ^7+\right.\right.\\
   & &\left.\left.154832 \lambda_i ^6+2705144 \lambda_i ^5+27068168 \lambda_i
   ^4+140926760 \lambda_i ^3+\right.\right.\\
   & &\left.\left.253029340 \lambda_i ^2-493864887 \lambda_i -1682001090\right)\right)
\end{eqnarray*}

Taking $\tilde{\mu}=\mu$, $\lambda_i=\frac{5}{2}$\quad, b=0.01 and solving the above equation, we get
\begin{eqnarray*}
 \mu_{min}=1.89037
\end{eqnarray*}
where we have discarded the negative value of $\mu_{min}$.

%--------------------------------------------------------------------
\end{document}